\documentclass{article}

\usepackage[nonatbib, preprint]{neurips_2023_gaied}

\usepackage[nonatbib]{neurips_2023_gaied}

\usepackage[utf8]{inputenc} 
\usepackage[T1]{fontenc}    
\usepackage{hyperref}       
\usepackage{url}            
\usepackage{booktabs}       
\usepackage{amsfonts}       
\usepackage{nicefrac}       
\usepackage{microtype}      
\usepackage[most]{tcolorbox}
\usepackage{xcolor}         
\usepackage{biblatex}
\title{Generative AI in the Classroom: \\ Can Students Remain Active Learners?}

\author{%
Rania Abdelghani$^{1,2}$ \quad H\'el\`ene Sauz\'eon$^{1}$ \quad Pierre-Yves Oudeyer$^{1}$ \\ \\
$^1$INRIA research center University of Bordeaux, Talence, France \\ $^2$EvidenceB, Paris, France\\
}

\begin{document}

\maketitle

\begin{abstract}
Generative Artificial Intelligence (GAI) can be seen as a double-edged weapon in education. Indeed, it may provide personalized, interactive and empowering pedagogical sequences that could favor students' intrinsic motivation, active engagement and help them have more control over their learning. But at the same time, other GAI properties such as the lack of uncertainty signalling even in cases of failure (particularly with Large Language Models (LLMs)) could lead to opposite effects, e.g. over-estimation of one's own competencies, passiveness, loss of curious and critical-thinking sense, etc.

These negative effects are due in particular to the lack of a pedagogical stance in these models' behaviors. Indeed, as opposed to standard pedagogical activities, GAI systems are often designed to answers users' inquiries easily and conveniently, without asking them to make an effort, and  without focusing on their learning process and/or outcomes.

This article starts by outlining some of these opportunities and challenges surrounding the use of GAI in education, with a focus on the effects on students' active learning strategies and related metacognitive skills. Then, we present a framework for introducing pedagogical transparency in GAI-based educational applications. This framework presents 1) training methods to include pedagogical principles in the models, 2) methods to ensure controlled and pedagogically-relevant interactions when designing activities with GAI and 3) educational methods enabling students to acquire the relevant skills to properly benefit from the use of GAI in their learning activities (meta-cognitive skills, GAI litteracy).



\end{abstract}

\section{Introduction}
The rise of generative artificial intelligence (GAI) represents an unprecedented opportunity for dealing with the multiple challenges facing the educational field today. Indeed, in a context where students are having more and more diverse needs, and educational establishments suffering from more and more teachers shortages, GAI can play a crucial role. One of the most straightforward implications is using LLMs to help educators create pedagogical content, personalized learning sequences and feedback, attractive and collaborative interactions~\cite{kasneci2023}, etc. Beyond helping teachers be more efficient, these new opportunities can also facilitate students' intrinsic motivation and active engagement during learning~\cite{jones2009}. This is indeed a strong argument in favour of using LLMs in education since such factors are shown to be fundamental for an efficient and enjoyable learning experience~\cite{borah2021, divjak2011, lin2017}, ... .

But although these seem as undeniable advantages, it is important to remember one of the core features of LLMs: they are designed to carry out users' instructions without the need -or with a minimal need- for a specific context~\cite{brown2020, zhou2023, wei2021}. This property, the forte of LLMs, needs to be taken with a lot of caution in education. Indeed, in order to be efficient learners, it is very well documented that students need to be able to frame informative and precise inquiries~\cite{oudeyer2016} and receive answers accordingly. In other words, in order for these models to be efficient learning companions, they should push students towards being active and in control of their learning by delivering information following specific pedagogical considerations (format, content, structure, etc). However, LLMs today are still lacking these considerations. For example, Macina et al.~\cite{macina2023} highlight the fact that, even when giving accurate answers in mathematics problems, these solutions are given very quickly by the LLM (ChatGPT), which doesn't give the opportunity for students to think, adapt, etc. And although using training techniques such as Reinforcement Learning from Human feedback (RLHF) can help move forward towards this goal, this method still presents several limitations mainly due to the absence of pedagogical stance in humans' feedback, the quality of their feedback (errors, wrong intentions, biases, etc), differences in the points of view between different evaluators, etc~\cite{casper2023}.
This can also be seen as a part of a more general challenge with LLMs which is their high controllability (i.e. the possibility to easily condition and sway their outputs towards a wanted behavior). This could be particularly risky in education as it means that, intentionally or not, students can make LLMs behave in opposite directions of pedagogical goals and methods, very easily. For example, research such as~\cite{rastogi2023, hartmann2023, kovavc2023} highlights the idea that LLMs may be viewed as 'agents' having 'ideas', 'positions', or 'ethical values' is misleading. This makes it unclear whether the current training approaches can align LLMs' behaviors with specific pedagogical policies and goals. 

\begin{figure}[ht]
  \centering
  \includegraphics[width=1\linewidth]{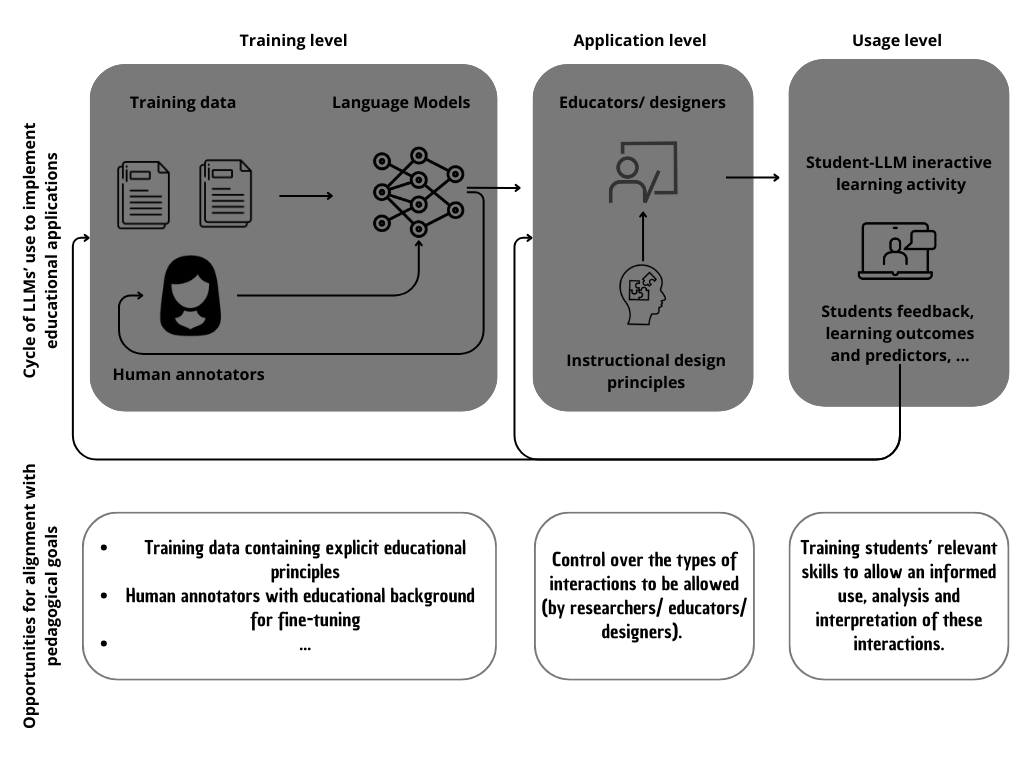}
  \caption{\textbf{Life-cycle of LLMs usage in educational applications and opportunities for alignment with pedagogical goals.}}
  \label{fig:illustration}
\end{figure}

If such challenges are not addressed when designing GAI-based pedagogical activities, they can be conducive to poor states of learning awareness (i.e. inability to understand and control the learning process), which is an important asset to favor motivation, activeness and to fully benefit from the learning activity. Several levels can be considered to face these challenges. First, on the training level: by grounding pedagogical methods and goals through the specific training data, alignment, etc. Second, on the application/ implementation level: by making informed decisions about 
how to use GAI to implement sufficient pedagogical activities (e.g. types of interactions allowed, controllability level allowed, etc). And third, on the usage level: by helping students develop the relevant skills that could help them be aware of and overcome the challenges of using GAI. These levels are summarized in Figure~\ref{fig:illustration}.

In this context, the current article proposes to start by over-viewing the opportunities and challenges that surround students' use of GAI during learning. In doing so, we focus on LLMs and their effects on active learning strategies and on predictors of learning such as critical thinking, metacognition, etc. Once laid out, we try to discuss some possibilities to face these challenges on the usage level, from an educational psychology point of view: how could we prepare students, especially the younger ones, to have informed and efficient use of such tools during learning? We focus on the role of metacognitive training and GAI literacy.

This work tries to suggest future research tracks that need to be explored in order to prepare for a safer and more efficient use of GAI in education. By doing so, it calls for joint effort from actors in the AI community, educational psychology research, educators and EdTech designers.

\section{How can the use of GAI affect students’ active learning ?}
Adopting active learning strategies refers to a student's ability to stay engaged during learning by exploring the environment and 'working' for information rather than receiving it passively. This includes exhibiting curiosity, epistemic vigilance and critical thinking, metacognitive self-regulation, etc.

Immediately, we can think of several GAI properties that could have positive or negative effects on such dimensions. We discuss these properties in the next paragraphs.


\subsection{Opportunities for active learning}
One of the most straightforward applications of GAI ---and particularly LLMs-- in education, is the idea of taking advantage of their strong linguistic abilities to implement incentive interactions that can push students towards adopting more active learning strategies~\cite{kasneci2023}. This can be done by designing specific interactions to trigger deeper and higher-level thinking capabilities. Some previous work has indeed showed some promising results concerning such applications. For example, work in~\cite{abdelghani2023a} showed that GPT-3 can be efficient in prompting students' curiosity during reading-comprehension tasks. It helped them ask deeper-thinking and more divergent questions by proposing hints that help them identify their own uncertainties. Moreover, LLMs could constitute a strong tool to improve critical thinking skills. Indeed, such skills are shown to be supported by adding opportunities for dialogue during learning, exposure to authentic or situated problems/ examples, etc~\cite{abrami2015}. All of these strategies could be much more easily implemented in the classroom by using LLMs, given their interactive and properties and access to huge amounts of scenarios and examples. Furthermore, LLMs can also facilitate the implementation of educational activities where children play active role, such as collaborators or teachers. Such roles can help them gain evaluative and monitoring metacognitive skills since they need to have a deep understanding about a given knowledge component in order to be able to teach it and/or use it to communicate with peers~\cite{Loewenstein94}.

On another hand, one strong advantage of LLMs is their ability to introduce personalized pedagogical guidance during learning~\cite{biswas2023, zhai2022}. This could have several functions: informative feedback to assess the student's cognitive performance, metacognitive prompting to propose useful learning strategies to adopt, etc. Implementing such behaviors could be of a great benefit for students, both on the cognitive and metacognitive levels, as it could help them have the relevant information to 1) make accurate mental models of the pedagogical concepts at hand. And 2) to auto-regulate their learning and be more efficient in their strategies to reach their goals.

Overall, LLMs can present several important opportunities to be efficient learning companions. However, it is clear that in order to reach these opportunities, careful design work needs to be done in order to implement  specific and complex interactions that could bypass the inherent lack of pedagogical stance in these models' behaviors. But what about allowing uncontrolled, spontaneous interactions ?

\subsection{Challenges for active learning}
\subsection{GAI challenges the ability to work for information}
One of the very relevant challenges of allowing uncontrolled interactions with LLMs is the risk of students becoming overly-relying on receiving information with minimal-to-none effort, and believing that solving any type of problem can be easily accessible without the need for any specific previous knowledge~\cite{kasneci2023}. This could lead to passiveness and undermine the intrinsic motivation to conduct investigations and active information-search. Furthermore, this apparent convenience and facility to obtain information with LLMs could lead to a negative learning control~\cite{bjork2013, baidoo2023}, meaning that students could start linking cognitive success with these powerful tools (LLMs), rather than with active learning strategies such as question-asking, exploration, etc. These are indeed crucial points to consider given their significant links with learning outcomes~\cite{gruber2019, oudeyer2016}.

\subsection{GAI challenges the development of critical thinking self-reflective behaviors}
Another LLMs' property that could lead to the loss of learning control is the lack of uncertainty signalling and the continuous confidence they exhibit even in cases of failure. This behavior means that students are not given the opportunity to judge in depth the quality of the information delivered by the model. This could result in two major problems:
\begin{enumerate}
    \item \textbf{An over-dimensioned representation of the knowledgeability of LLMs.} This can lead to systematically accepting their behaviors without much analytic or critical thinking~\cite{kidd2023}. It is particularly sensitive in education since evidence shows that individuals tend to track the knowledgeability of agents and use this signal to form their own beliefs and information-search behaviors~\cite{sabbagh2001, heider1944, orticio2023}. This means that, when mis-representing an agent's knowledgeability, individuals can form wrong and/or biased mental models of the knowledge component at hand. Once transferred to them, these misinformation, biases, etc cannot be easily corrected ~\cite{thompson2021}.
    \item \textbf{An over-estimation of students' own knowledge.} The stream of affirmation during interactions with LLMs means that students will have very little chance to reflect on their knowledge states and analyze their progress, strengths and weaknesses for the task. Having an over-estimation of one's own competencies is one of the first brakes that can keep individuals from engaging in exploratory and active information-search behaviors~\cite{Loewenstein94}.
\end{enumerate}
These are major  and realistic challenges given that LLMs are subject to several failures in their behaviors, such as unsubstantiated content, hallucinations, incomplete information, ambiguity, cognitive and/or social biases, etc~\cite{chen2023, jones2022}. In this line, empirical studies such as~\cite{hill2023} show indeed that even university students fail to detect failures in ChatGPT's answers and tended to use them in exams without analyzing/ correcting them. Furthermore, work in~\cite{chen2023} shows that LLM-generated misinformation can be harder to detect for humans, compared to human-generated misinformation. This suggests that LLMs can potentially cause more harm and expose students to adopting more wrong/ biased information compared to standard learning settings.

We believe that all of these challenges we raise, and more, are essentially coming from one higher-level barrier: as opposed to the primary objectives of standard pedagogical activities, LLMs are not designed ---and do not behave in a way--- to carry out pedagogical goals and support users' learning. Even with training methods such as RLHF that are put in place to align LLMs with human goals, it is still not clear what is the pedagogical intent (if any) behind the training of these models. Indeed, RLHF involves human evaluators who can be of various backgrounds, opinions, goals and expertise.

\section{How can we promote active learning in a GAI era ?}
As we mention in the introduction and illustrate in Figure~\ref{fig:illustration}, facing active learning-related challenges can be initiated by different actors and on different levels:
\begin{itemize}
    \item \textbf{Role of the AI community on the training level:} consists of refining the training policies and content of the models, in order to acheive a pedagogical methods/ goals-grounding. Concretely, this could be done by training the model on more educational and learning principles data, recruiting teachers as annotators during the RLHF process, etc.
    \item \textbf{Role of educators/ EdTech designers on the application/ implementation level:} consists of making informed decisions when using LLMs to design educational activities (e.g. by ensuring enough control, pedagogically-oriented interactions, etc).
    \item \textbf{Role of researchers on the usage level:} consists of preparing children to be aware of GAI's challenges and be able to use it in an informed and positive way. It relies on supporting them to develop the relevant knowledge and skills that could help them stay active and critical when using this particular technology.
\end{itemize}

In this next section, we focus on the third point, i.e. the usage level. We argue that in order to stay active and on control of their learning while using GAI, students should be able to understand the latter's behaviors rationale and have strong metacognitive skills.


\subsection{Role of GAI literacy}
As already discussed, one of the biggest challenges of using GAI is the misleading over-hype around its agency and knowledgeability. To have a more realistic representation of these tools, we think that it is important for students to see their real potential, strengths and weaknesses and understand the motivation and rationale behind their design. For example, explaining that these models rely heavily on statistical methods to predict their next behaviors, can help students see the difference with how biological brains work~\cite{brown2020}. At its turn, this can help have a more accurate representation of these models' power: they are likely to show several problems related to cognitive reasoning, senses of the physical world, logic, etc that are difficult for users to disentangle~\cite{jones2022}.

We also think that it is important to raise students' attention to the different failures that GAI ---and LLMs specifically--- can generate. Indeed, problems such as outdated information, unsubstantiated content, incomplete information, lack of context, complete fabrication,... can easily occur when interacting with LLMs~\cite{chen2023, novelli2023}. Furthermore, students need also to be aware of the existence of cognitive and/or social bias in these models' behaviors~\cite{casper2023} that can emerge from the initial training datasets, the human annotators implicated in the RLHF process, etc~\cite{bender2021dangers}. Understanding such specific limitations can push students to be more cautious, analytic and critic of GAI's behaviors and therefore, to think deeper before adopting them. Such explanations can take the form of a series of educational videos for example. Several contents introducing GAI to the general public are currently being developed~\cite{torres2023} and can be adapted to students' levels.

On another hand, it is important to remember that the ability to conduct efficient information-search strategies in rich online environments depends on students' ability to form specific and goal-directed inquiries~\cite{rouet2003}. In GAI environments, this translates to the importance of using specific and clear prompting strategies. For example, work in~\cite{wei2022} shows that Chain-Of-Though prompting leads LLMs to a better performance in complex reasoning tasks. For this reason, we think that teaching students efficient prompting strategies is important, not only because it can enhance GAI's behaviors, but also because it can help students understand their tasks better. Indeed, an efficient prompting process requires individuals to think about how to explain the task well and introduce its context clearly to the system, decompose it into several steps if necessary, specify the nature of outcome that is wanted, etc. This means that teaching the good strategies of interacting with GAI can also lead to better metacognitive states and awareness about the learning process and goals.

Finally, we think it is also very relevant to propose such trainings to teachers and the educational designers/ developers in order to help them have clearer views about the potential of GAI and how to use it appropriately in order to implement efficient pedagogical activities that can support students' active and meaningful learning. 

\subsection{Role of metacognitive training}
Although very important, GAI literacy may not be sufficient for nurturing an informed use of these systems in education. For example, work in~\cite{theophilou2023} shows that interventions targeting prompting strategies led to enhanced states of trust in the system by students. Although this can be seen as a positive result, it can also mean that students are systematically trusting the system's behaviors as long as they're using the prompting strategies they learnt are 'correct'. This could lead to the same challenges we pointed above, i.e. over-reliance on GAI, loss of critical/ analytic instincts, etc.

To face this, we think that GAI literacy should be coupled with the training of higher-level skills such as metacognition, critical-thinking or intellectual vigilance, that are key components to active control during learning~\cite{livingston2003}. Indeed, metacognition is argued to be crucial to engage in active knowledge-acquisition behaviors: Loewnstein~\cite{Loewenstein94} explains that curiosity-driven information-search is primarily motivated by the individual's ability to detect a knowledge gap in their own knowledge. This step mobilizes the metacognitive ability to evaluate one's own knowledge and decide about the next relevant cognitive behaviors to adopt. This means that if we want students to be more at ease with asking further information when they feel surprised or/and uncertain about GAI's behavior, supporting them to develop evaluative metacognitive skills can be an efficient strategy. Indeed, preliminary work in~\cite{abdelghani2023b} shows that training specific metacognitive skills resulted in more epistemic curiosity behaviors in 9-to-11 yo children.



Moreover, critical thinking skills also depend on metacognition as they require the individual to be able to pursue their goals through self-directed search and interrogation of knowledge, link the information they find with their prior knowledge, hypotheses, etc~\cite{murayama2019}. Thus, it becomes imperative to introduce students to these skills, enabling them to discern the need for corroborative evidence, both supporting and challenging GAI-generated information, prior to acceptance. Such behaviors can help students maintain a purposeful learning trajectory when using GAI for resolving specific cognitive tasks.

It is to be noted that several studies have shown that metacognition can be trained using specific methods and prompting approaches such as inciting self-reflection, evaluating newly-acquired information, etc, ~\cite{abdelghani2022, abdelghani2023b, alaimi2020}, ... . However, to the best of our knowledge, no studies have begun exploring applying such metacognitive-prompting approaches in GAI learning environments where interactions are more complex and challenging. For this reason we think that it is important to work on adapting these approaches to the properties and challenges of GAI and design new interventions.

\section{Conclusion}
Generative AI presents huge opportunities for enhancing the teaching and learning experiences and outcomes. It has the potential to address several challenges facing the educational sector today, from teachers' shortages to taking charge of individual differences in students. However, using these systems in the classrooms still raises a lot of questions, especially concerning the lack of pedagogical intentions in these systems, which can lead to interactions that foster passiveness and loss of learning control. 


With this paper, we try to lay out some of the opportunities and challenges associated with using GAI in the classroom, with a focus on how it may impact student's active learning strategies and metacognition. From there, we discuss the roles of the different stakeholders in facing these challenges, from the AI community, to educational psychologists and educators. This joint effort can indeed lead to having more pedagogically-balanced GAI and better-skilled students that can use these tools to learn in an active and responsible way.


\begin{ack}
This work benefited from funding from the EvidenceB company (CIFRE Grant for Rania Abdelghani) and from ANR AI Chair DeepCuriosity ANR-19-CHIA-0004. Pierre-Yves Oudeyer is also scientific consultant at EvidenceB. 
\end{ack}

\printbibliography

@article{kasneci2023,
	title        = {ChatGPT for good? On opportunities and challenges of large language models for education},
	author       = {Kasneci, Enkelejda and Se{\ss}ler, Kathrin and K{\"u}chemann, Stefan and Bannert, Maria and Dementieva, Daryna and Fischer, Frank and Gasser, Urs and Groh, Georg and G{\"u}nnemann, Stephan and H{\"u}llermeier, Eyke and others},
	year         = 2023,
	journal      = {Learning and individual differences},
	publisher    = {Elsevier},
	volume       = 103,
	pages        = 102274
}

@inproceedings{bender2021dangers,
	title        = {On the dangers of stochastic parrots: Can language models be too big?},
	author       = {Bender, Emily M and Gebru, Timnit and McMillan-Major, Angelina and Shmitchell, Shmargaret},
	year         = 2021,
	booktitle    = {Proceedings of the 2021 ACM conference on fairness, accountability, and transparency},
	pages        = {610--623}
}

@article{jones2009,
	title        = {Motivating students to engage in learning: the MUSIC model of academic motivation.},
	author       = {Jones, Brett D},
	year         = 2009,
	journal      = {International Journal of Teaching and Learning in Higher Education},
	publisher    = {ERIC},
	volume       = 21,
	number       = 2,
	pages        = {272--285}
}

@article{lin2017,
	title        = {A study of the effects of digital learning on learning motivation and learning outcome},
	author       = {Lin, Ming-Hung and Chen, Huang-Cheng and Liu, Kuang-Sheng},
	year         = 2017,
	journal      = {Eurasia Journal of Mathematics, Science and Technology Education},
	publisher    = {Modestum},
	volume       = 13,
	number       = 7,
	pages        = {3553--3564}
}

@article{divjak2011,
	title        = {The impact of game-based learning on the achievement of learning goals and motivation for learning mathematics-literature review},
	author       = {Divjak, Bla{\v{z}}enka and Tomi{\'c}, Damir},
	year         = 2011,
	journal      = {Journal of information and organizational sciences},
	publisher    = {Fakultet organizacije i informatike Sveu{\v{c}}ili{\v{s}}ta u Zagrebu},
	volume       = 35,
	number       = 1,
	pages        = {15--30}
}

@article{borah2021,
	title        = {Motivation in learning},
	author       = {Borah, Mayuri},
	year         = 2021,
	journal      = {Journal of Critical Reviews},
	volume       = 8,
	number       = 2,
	pages        = {550--552}
}

@article{zhou2023,
	title        = {A comprehensive survey on pretrained foundation models: A history from bert to chatgpt},
	author       = {Zhou, Ce and Li, Qian and Li, Chen and Yu, Jun and Liu, Yixin and Wang, Guangjing and Zhang, Kai and Ji, Cheng and Yan, Qiben and He, Lifang and others},
	year         = 2023,
	journal      = {arXiv preprint arXiv:2302.09419}
}

@article{wei2021,
	title        = {Finetuned language models are zero-shot learners},
	author       = {Wei, Jason and Bosma, Maarten and Zhao, Vincent Y and Guu, Kelvin and Yu, Adams Wei and Lester, Brian and Du, Nan and Dai, Andrew M and Le, Quoc V},
	year         = 2021,
	journal      = {arXiv preprint arXiv:2109.01652}
}

@article{brown2020,
	title        = {Language models are few-shot learners},
	author       = {Brown, Tom and Mann, Benjamin and Ryder, Nick and Subbiah, Melanie and Kaplan, Jared D and Dhariwal, Prafulla and Neelakantan, Arvind and Shyam, Pranav and Sastry, Girish and Askell, Amanda and others},
	year         = 2020,
	journal      = {Advances in neural information processing systems},
	volume       = 33,
	pages        = {1877--1901}
}

@misc{casper2023,
	title        = {Open Problems and Fundamental Limitations of Reinforcement Learning from Human Feedback},
	author       = {Stephen Casper and Xander Davies and Claudia Shi and Thomas Krendl Gilbert and J\'er\'emy Scheurer and Javier Rando and Rachel Freedman and Tomasz Korbak and David Lindner and Pedro Freire and Tony Wang and Samuel Marks and Charbel-Raphaël Segerie and Micah Carroll and Andi Peng and Phillip Christoffersen and Mehul Damani and Stewart Slocum and Usman Anwar and Anand Siththaranjan and Max Nadeau and Eric J. Michaud and Jacob Pfau and Dmitrii Krasheninnikov and Xin Chen and Lauro Langosco and Peter Hase and Erdem Bıyık and Anca Dragan and David Krueger and Dorsa Sadigh and Dylan Hadfield-Menell},
	year         = 2023,
	eprint       = {2307.15217},
	archiveprefix = {arXiv},
	primaryclass = {cs.AI}
}

@article{theophilou2023,
	title        = {Learning to Prompt in the Classroom to Understand AI Limits: A pilot study},
	author       = {Theophilou, Emily and Koyuturk, Cansu and Yavari, Mona and Bursic, Sathya and Donabauer, Gregor and Telari, Alessia and Testa, Alessia and Boiano, Raffaele and Hernandez-Leo, Davinia and Ruskov, Martin and others},
	year         = 2023,
	journal      = {arXiv preprint arXiv:2307.01540}
}

@article{kidd2023,
	title        = {How AI can distort human beliefs},
	author       = {Kidd, Celeste and Birhane, Abeba},
	year         = 2023,
	journal      = {Science},
	publisher    = {American Association for the Advancement of Science},
	volume       = 380,
	number       = 6651,
	pages        = {1222--1223}
}

@article{hill2023,
	title        = {Taking the help or going alone: ChatGPT and class assignments},
	author       = {Hill, Brian},
	year         = 2023,
	journal      = {HEC Paris Research Paper Forthcoming}
}

@article{gruber2019,
	title        = {Curiosity and learning: a neuroscientific perspective},
	author       = {Gruber, Matthias J and Valji, Ashanti and Ranganath, Charan},
	year         = 2019,
	publisher    = {Cambridge University Press}
}

@article{abdelghani2022,
	title        = {Conversational agents for fostering curiosity-driven learning in children},
	author       = {Abdelghani, Rania and Oudeyer, Pierre-Yves and Law, Edith and de Vulpilli{\`e}res, Catherine and Sauz{\'e}on, H{\'e}l{\`e}ne},
	year         = 2022,
	journal      = {International Journal of Human-Computer Studies},
	publisher    = {Elsevier},
	volume       = 167,
	pages        = 102887
}

@article{abdelghani2023a,
	title        = {Gpt-3-driven pedagogical agents to train children’s curious question-asking skills},
	author       = {Abdelghani, Rania and Wang, Yen-Hsiang and Yuan, Xingdi and Wang, Tong and Lucas, Pauline and Sauz{\'e}on, H{\'e}l{\`e}ne and Oudeyer, Pierre-Yves},
	year         = 2023,
	journal      = {International Journal of Artificial Intelligence in Education},
	publisher    = {Springer},
	pages        = {1--36}
}

@inproceedings{abdelghani2023b,
	title        = {Interactive environments for training children’s curiosity through the practice of metacognitive skills: a pilot study},
	author       = {Abdelghani, Rania and Law, Edith and Desvaux, Chlo{\'e} and Oudeyer, Pierre-Yves and Sauz{\'e}on, H{\'e}l{\`e}ne},
	year         = 2023,
	booktitle    = {Proceedings of the 22nd Annual ACM Interaction Design and Children Conference},
	pages        = {495--501}
}

@article{baidoo2023,
	title        = {Education in the era of generative artificial intelligence (AI): Understanding the potential benefits of ChatGPT in promoting teaching and learning},
	author       = {Baidoo-Anu, David and Ansah, Leticia Owusu},
	year         = 2023,
	journal      = {Journal of AI},
	publisher    = {{\.I}zmir Academy Association},
	volume       = 7,
	number       = 1,
	pages        = {52--62}
}

@inproceedings{alaimi2020,
	title        = {Pedagogical agents for fostering question-asking skills in children},
	author       = {Alaimi, Mehdi and Law, Edith and Pantasdo, Kevin Daniel and Oudeyer, Pierre-Yves and Sauzeon, H{\'e}l{\`e}ne},
	year         = 2020,
	booktitle    = {Proceedings of the 2020 CHI Conference on Human Factors in Computing Systems},
	pages        = {1--13}
}

@article{bjork2013,
	title        = {Self-regulated learning: Beliefs, techniques, and illusions},
	author       = {Bjork, Robert A and Dunlosky, John and Kornell, Nate},
	year         = 2013,
	journal      = {Annual review of psychology},
	publisher    = {Annual Reviews},
	volume       = 64,
	pages        = {417--444}
}

@article{sabbagh2001,
	title        = {Learning words from knowledgeable versus ignorant speakers: Links between preschoolers' theory of mind and semantic development},
	author       = {Sabbagh, Mark A and Baldwin, Dare A},
	year         = 2001,
	journal      = {Child development},
	publisher    = {Wiley Online Library},
	volume       = 72,
	number       = 4,
	pages        = {1054--1070}
}

@article{thompson2021,
	title        = {Human biases limit cumulative innovation},
	author       = {Thompson, Bill and Griffiths, Thomas L},
	year         = 2021,
	journal      = {Proceedings of the Royal Society B},
	publisher    = {The Royal Society},
	volume       = 288,
	number       = 1946,
	pages        = 20202752
}

@article{heider1944,
	title        = {An experimental study of apparent behavior},
	author       = {Heider, Fritz and Simmel, Marianne},
	year         = 1944,
	journal      = {The American journal of psychology},
	publisher    = {JSTOR},
	volume       = 57,
	number       = 2,
	pages        = {243--259}
}

@article{Loewenstein94,
	title        = {The Psychology of Curiosity: A Review and Reinterpretation},
	author       = {G. Loewenstein},
	year         = 1994,
	month        = jul,
	journal      = {Psychological Bulletin},
	volume       = 116,
	pages        = {75--98},
	doi          = {10.1037/0033-2909.116.1.75}
}

@article{hartmann2023,
	title        = {The political ideology of conversational AI: Converging evidence on ChatGPT's pro-environmental, left-libertarian orientation},
	author       = {Hartmann, Jochen and Schwenzow, Jasper and Witte, Maximilian},
	year         = 2023,
	journal      = {arXiv preprint arXiv:2301.01768}
}

@inproceedings{rastogi2023,
	title        = {Supporting human-ai collaboration in auditing llms with llms},
	author       = {Rastogi, Charvi and Tulio Ribeiro, Marco and King, Nicholas and Nori, Harsha and Amershi, Saleema},
	year         = 2023,
	booktitle    = {Proceedings of the 2023 AAAI/ACM Conference on AI, Ethics, and Society},
	pages        = {913--926}
}

@article{kovavc2023,
	title        = {Large Language Models as Superpositions of Cultural Perspectives},
	author       = {Kova{\v{c}}, Grgur and Sawayama, Masataka and Portelas, R{\'e}my and Colas, C{\'e}dric and Dominey, Peter Ford and Oudeyer, Pierre-Yves},
	year         = 2023,
	journal      = {arXiv preprint arXiv:2307.07870}
}

@article{jones2022,
	title        = {Capturing failures of large language models via human cognitive biases},
	author       = {Jones, Erik and Steinhardt, Jacob},
	year         = 2022,
	journal      = {Advances in Neural Information Processing Systems},
	volume       = 35,
	pages        = {11785--11799}
}

@article{novelli2023,
	title        = {Taking AI Risks Seriously: a Proposal for the AI Act},
	author       = {Novelli, Claudio and Casolari, Federico and Rotolo, Antonino and Taddeo, Mariarosaria and Floridi, Luciano},
	year         = 2023,
	journal      = {Available at SSRN 4447964}
}

@article{murayama2019,
	title        = {Process account of curiosity and interest: A reward-learning perspective},
	author       = {Murayama, Kou and FitzGibbon, Lily and Sakaki, Michiko},
	year         = 2019,
	journal      = {Educational Psychology Review},
	publisher    = {Springer},
	volume       = 31,
	pages        = {875--895}
}

@article{livingston2003,
	title        = {Metacognition: An Overview.},
	author       = {Livingston, Jennifer A},
	year         = 2003,
	publisher    = {ERIC}
}

@misc{torres2023,
	title        = {ChatGPT en 5mn: une s\'erie p\'edagogique pour le grand public},
	author       = {Torres-Leguet, Alexandre and Romac, Cl\'e ment and Carta, Thomas and Oudeyer, Pierre-Yves},
	year         = 2023,
	url          = {https://developmentalsystems.org/chatgpt_en_5_minutes/}
}

@article{oudeyer2016,
	title        = {Intrinsic motivation, curiosity, and learning: Theory and applications in educational technologies},
	author       = {Oudeyer, P-Y and Gottlieb, Jacqueline and Lopes, Manuel},
	year         = 2016,
	journal      = {Progress in brain research},
	publisher    = {Elsevier},
	volume       = 229,
	pages        = {257--284}
}

@article{abrami2015,
	title        = {Strategies for teaching students to think critically: A meta-analysis},
	author       = {Abrami, Philip C and Bernard, Robert M and Borokhovski, Eugene and Waddington, David I and Wade, C Anne and Persson, Tonje},
	year         = 2015,
	journal      = {Review of educational research},
	publisher    = {Sage Publications Sage CA: Los Angeles, CA},
	volume       = 85,
	number       = 2,
	pages        = {275--314}
}

@article{biswas2023,
	title        = {Role of Chat GPT in Education},
	author       = {Biswas, Som},
	year         = 2023,
	journal      = {Available at SSRN 4369981}
}

@article{zhai2022,
	title        = {ChatGPT user experience: Implications for education},
	author       = {Zhai, Xiaoming},
	year         = 2022,
	journal      = {Available at SSRN 4312418}
}

@inproceedings{orticio2023,
	title        = {Children flexibly adapt their evidentiary standards to their informational environment},
	author       = {Orticio, Evan and Meyer, Martin and Kidd, Celeste},
	year         = 2023,
	booktitle    = {Proceedings of the Annual Meeting of the Cognitive Science Society},
	volume       = 45,
	number       = 45
}

@article{chen2023,
	title        = {Can LLM-Generated Misinformation Be Detected?},
	author       = {Chen, Canyu and Shu, Kai},
	year         = 2023,
	journal      = {arXiv preprint arXiv:2309.13788}
}

@article{rouet2003,
	title        = {What was I looking for? The influence of task specificity and prior knowledge on students' search strategies in hypertext},
	author       = {Rouet, Jean-Fran{\c{c}}ois},
	year         = 2003,
	journal      = {Interacting with computers},
	publisher    = {OUP},
	volume       = 15,
	number       = 3,
	pages        = {409--428}
}

@article{wei2022,
	title        = {Chain-of-thought prompting elicits reasoning in large language models},
	author       = {Wei, Jason and Wang, Xuezhi and Schuurmans, Dale and Bosma, Maarten and Xia, Fei and Chi, Ed and Le, Quoc V and Zhou, Denny and others},
	year         = 2022,
	journal      = {Advances in Neural Information Processing Systems},
	volume       = 35,
	pages        = {24824--24837}
}

@misc{macina2023,
	title        = {MathDial: A Dialogue Tutoring Dataset with Rich Pedagogical Properties Grounded in Math Reasoning Problems},
	author       = {Jakub Macina and Nico Daheim and Sankalan Pal Chowdhury and Tanmay Sinha and Manu Kapur and Iryna Gurevych and Mrinmaya Sachan},
	year         = 2023,
	eprint       = {2305.14536},
	archiveprefix = {arXiv},
	primaryclass = {cs.CL}
}
\end{document}